\documentclass[pre,aps,amsmath,twocolumn,amssymb,superscriptaddress]{revtex4-1}%
\usepackage{amsmath}
\usepackage{amsfonts}
\usepackage{amssymb}
\usepackage{graphicx}
\usepackage{color}
\usepackage{version}
\usepackage{float,subfigure}
\newcommand{\be}{\begin{equation}}
\newcommand{\ee}{\end{equation}}

\begin{document}
\title{Effect of confinement on dense packings of rigid frictionless spheres and polyhedra}
\author{Jean-Fran\c cois Camenen}
\email{jfcamenen@gmail.com}
\affiliation{LUNAM Universit\'e, IFSTTAR, site de Nantes, Route de Bouaye CS4 44344 Bouguenais Cedex, France}
\author{Yannick Descantes}
\email{yannick.descantes@ifsttar.fr}
\affiliation{LUNAM Universit\'e, IFSTTAR, site de Nantes, Route de Bouaye CS4 44344 Bouguenais Cedex, France}
\author{Patrick Richard}
\email{patrick.richard@ifsttar.fr}
\affiliation{LUNAM Universit\'e, IFSTTAR, site de Nantes, Route de Bouaye CS4 44344 Bouguenais Cedex, France}
\affiliation{Institut Physique de Rennes, Universit\'e de Rennes I, UMR CNRS 6251, 35042 Rennes, France}

\date{\today}

\begin{abstract}
We study numerically the influence of confinement on the solid fraction  and on the structure of three-dimensional random close packed (RCP) granular materials subject to gravity. The effects of grain shape (spherical or polyhedral), material polydispersity and confining wall friction on this dependence are investigated. 
{In agreement with a {simple} geometrical model,} the solid fraction is found to decrease linearly for increasing confinement no matter the grain shape. Furthermore, this decrease remains valid for bidisperse sphere packings although the gradient seems to reduce significantly when the proportion of small particles reaches $40\%$ by volume. 
The confinement effect on the coordination number is also captured by an extension of the aforementioned model. 
\end{abstract}

\maketitle
\section{Introduction}\label{sec:intro}
Granular materials are well known for their wide range of fascinating properties. Their theoretical description is difficult for many reasons. One of them is the importance of the local arrangement of grains within the material on its macroscopic behavior.
Real granular systems have boundaries, but for the sake of simplicity, scientists often neglect them~(see for example \cite{Silbert_PRE_2002,Agnolin_PRE_2007}) and the system is then considered as infinite. 
This {assumption} is not always justified since the presence of boundaries modifies the {system} local arrangement in their vicinity. Moreover, due to the intrinsic steric hindrance of granular materials those structure modifications often propagate over distances in the order of several grain sizes. As a consequence, the behavior of granular systems may be strongly influenced by the presence of sidewalls even if the confinement length is large compared to the grain size.\\
The crucial role of confinement on system properties has been pointed out in many works dealing with gravity driven granular flows~\cite{Taberlet2003,Taberlet2004b,Jop2005,Taberlet2006b,Richard2008,Taberlet2008}, granular segregation~\cite{Peirera2011}, structure and mechanics of granular packings~\cite{BenAim1968,Suzuki2008,Landry2003}, granular systems in narrow silos~\cite{Landry_powdertech_2004} or granular penetration by impact~\cite{Seguin2008}.
Those studies point out that two major physical properties can be influenced by the presence of sidewalls. First, they can induce friction that might be important in respect to the internal friction of the system~\cite{Richard2008} explaining the well-know Janssen effect~\cite{Vanel2000} or unexpectedly high angle values observed with confined granular heaps~\cite{Liu1991,Courrech2003} or confined chute flows~\cite{Taberlet2003}. Second, as mentioned above, they might also alter the {geometrical} structure of the system near the wall, where 
particles tend to form layers, giving rise to a fluctuating local solid fraction with distance from the wall~\cite{Suzuki2008} and affecting the properties of confined systems. 
Note that the effect of confinement is not limited to the vicinity of the walls but may propagate within the sample. This is more particularly the case for confined granular chute flows for which it has been shown that the good dimensionless number to quantify the sidewall effect is not the number of grains per unit of width between sidewalls but the ratio of the flow height to the gap between sidewalls~\cite{Taberlet2003}.\\
Here, we focus mainly on the geometric effect of the presence of sidewalls on quantities like the solid fraction and the coordination numbers.
Recently, Desmond and Weeks used numerical simulations to study the effect of confinement on binary atomic systems at the random-close-packing limit~\cite{Desmond2009}. 
Their numerical results agree with a simple geometrical model~\cite{Verman_Nature_1946,Brown_Nature_1946,Combe_PhD_2001} (hereafter called the geometrical model) which captures the evolution of the solid fraction of  random close packings of spheres with confinement. It is based on the following configuration: a packing of particles is confined between two parallel and flat walls separated by a gap $W$. It then assumes that such a confined system is made of two boundary layers (of thickness $h$) and a bulk region and that the solid fraction of the boundary layers, $\phi_{BL}$, is lower than that of the bulk 
{region}  $\phi_{bulk}$.
By writing the total solid fraction $\phi$ as the average of both the bulk 
{region} and boundary layers solid fractions (weighted by their relative thickness), the {geometrical} model predicts that the average solid fraction decreases 
{linearly with} 
$1/W$:
\be \label{eqn:weeks} \phi=\frac{W - 2h}{W}\phi_{bulk}+\frac{2h}{W}\phi_{BL}=\phi_{bulk}-\frac{C}{W},\ee
where $C=2h\left(\phi_{bulk}-\phi_{BL}\right)$. Note that this model can be easily adapted to other boundaries such as cylindrical ones~\cite{Desmond2009}. 
The three parameters of the {geometrical} model ($\phi_{bulk}$, $\phi_{BL}$ and $h$) probably depend on grain shape, packing polydispersity and confining wall properties.
Here we study the effect of confinement 
{on quasi-static dense} frictionless granular systems (i.e. grains interacting through hard core repulsion)
subjected to gravity. We test the validity of the 
geometrical model for such systems and study the aforementioned dependencies. Using numerical simulations, we investigate the actual effect of grain shape by comparing packings of spheres with packings of polyhedra. Furthermore, we assess the effect of packing 
{poly}dispersity by comparing monosized and binary packings. Besides, we check the effect of grain-wall friction. 
Eventually, we look into packing microstructure by studying the effect of confinement on the coordination number.\\

The paper is organized as follows. Section~\ref{sec:simmet} describes our simulations with details as well as the numerical simulation method used. After checking the state of packings in section~\ref{sec:state_packing}, section~\ref{sec:solid_fraction} investigates how the  solid fraction is modified by confinement and how these modifications are influenced by packing polydispersity, grain shape and confining wall friction. Then, we examine in section~\ref{sec:packing-microstructure} the modification of the packing microstructure  with confinement. Finally, in section~\ref{sec:conclusion} we summarize our results and present our conclusions.

\section{Simulation methodology}\label{sec:simmet}

\subsection{Geometry of grains}


The simulated system is a three-dimensional dense assembly of $n$ frictionless rigid grains of mass density $\rho$, interacting with each other through totally inelastic collisions.

\begin{figure}[htbp]
\begin{center}
\includegraphics*[width=1.0\columnwidth]{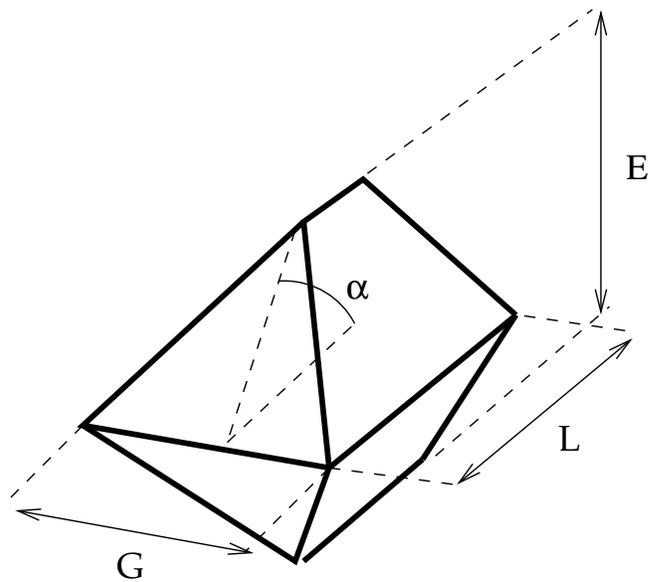} 
\caption{Pinacoid, a model polyhedra characterized by its length $L$, width $G$, height $E$ and angle $\alpha$.}  \label{pinacoid}
\end{center}
\end{figure}

Since 
grain shape may influence the behavior of granular materials~\cite{Atman_JPCM_2005,Ribiere2005,Ribiere2005c,Szarf2011}, two types of grains have been studied: spherical grains of average diameter $d$ and polyhedra of average characteristic dimension $d$. The polyhedra shape (Fig.~\ref{pinacoid}) is that of a \emph{pinacoid}, with eight vertices, fourteen edges and eight faces. This polyhedron has three symmetry planes and is determined by four parameters: length $L$, width $G$, height $E$ and angle $\alpha$. According to an extensive experimental study with various rock types reported
by~\cite{Tourenq82}, the pinacoid gives the best fit among simple geometries for an aggregate grain. In order to have similar aspect ratio for both grain shapes, the pinacoid dimensional
parameters were taken identical ($L=G=E$), with the characteristic dimension $d$ expressed as $d=L$. In addition, angle $\alpha$ was set to $60^\circ$. 
For each grain shape, two grain diameters (or characteristic dimensions) have been considered: large $d_L$ and small $d_S = d_L/2$. 

\subsection{Samples preparation}\label{subsec:samples_prep}

\begin{figure}[htbp]
\begin{center}
\includegraphics*[width=1.0\columnwidth]{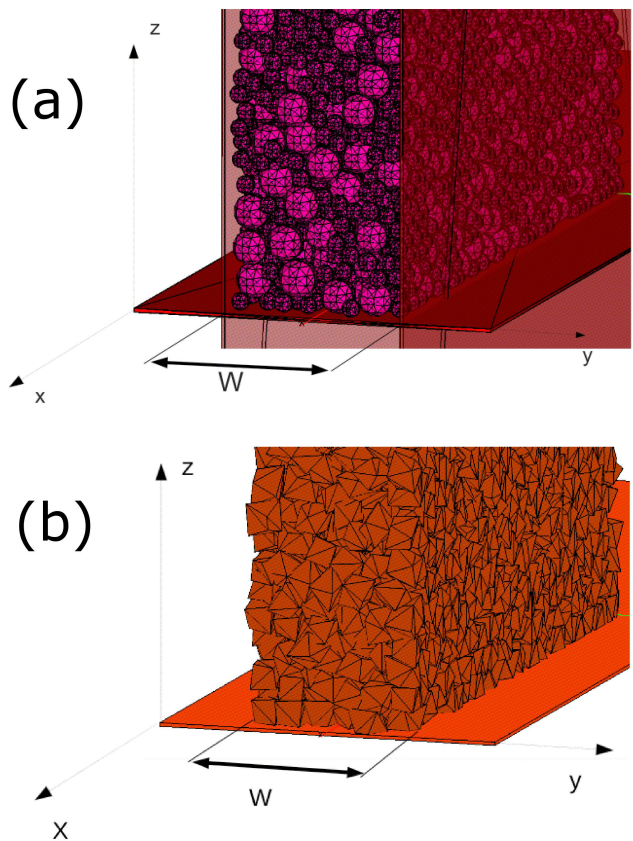}
\caption{(Color online) Typical 3D snapshots of packings made of polysized spheres (a) and pinacoids (b). The confinement is characterized by the gap 
{$W$} between sidewalls
. The direction of gravity is $-z$. }\label{packings} \end{center}
\end{figure}

The packing geometry is that of a parallelepiped (Fig.~\ref{packings}) of dimensions $L_x$ by $L_y$ by $L_z$. 
Periodic Boundary Conditions (PBC) 
{are} applied in the $x$ direction to simulate an infinitely long parallelepiped using a finite number of grains. 
The packing is confined 
{in} the $y$ direction between two fixed parallel walls 
{separated from each other by a $L_y=W$ large gap.}
In some cases, PBC are also applied along the $y$ axis to simulate unconfined reference state, with $W$ set to $20d_L$. 
The packing is supported on the $xy$-plane by a fixed frictionless bottom wall and delimited by a free surface at its top.

Grain samples are composed of various proportions of small and large grains having the same shape. In order to reduce the thickness of the crystallized layer commonly observed inside confined packings at the interface with smooth walls~\cite{murray_98,Teng_PRL_2003}, each population of grain size is randomly generated with a Gaussian distribution characterized by its mean $d$ and its variance $d^2/900$. Anyhow, for the sake of simplicity, packings made of a unique population of grains (either small or large) will be called "monodisperse", whereas packings made of small and large grains will be called "bidisperse" in the following. In the latter case, the proportions of small ($x_S$) and large ($x_L$) grains expressed as percentages by volume are of course linked through $x_S=100-x_L$.

Each sample is constructed layer-by-layer according to the following {geometrical} deposition protocol inspired by ~\cite{Laniel_PhD_2007}: spherical particles of a sufficient diameter to circumscribe the larger grains of the sample are sequentially dropped along $z$ in a parallelepiped box having the aforementioned geometry. Each particle stops on the free surface made of the underneath layer of particles (or on the bottom wall for the first layer of particles) and is further moved so that it lays on three particles chosen to locally minimize its altitude $z$. Finally, the sample actual grains are randomly substituted for those spherical particles. For polyhedra samples, a random orientation is further assigned to each pinacoid. Note that according to this protocol, some of the deposited grains may not be in contact with their neighbors depending on their size and shape.

\subsection{Initialization and solicitations} \label{subsec:init-solicit}

The system initialization is identical for spheres and pinacoids samples. The first step consists in geometrically depositing $n$ grains into a parallelepiped box, then PBC are substituted for the lateral walls of the parallelepiped box along the $x$ direction (along the $x$ and $y$ directions for biperiodic reference state). Finally, gravity $\vec g$ ($0$,$0$,$-g$) is applied in order to compact the sample.

\subsection{Contact dynamics method} \label{subsec:NSCD}

Discrete numerical simulations were performed using the contact
dynamics (CD) method~\cite{Moreau94,Moreau96}, which is specially
convenient for rigid grains. This method is based on implicit time
integration of the equations of motion with respect to generalized
non-smooth contact laws describing non-interpenetration and dry
friction between grains. This formulation unifies the description
of lasting contacts and collisions through the concept of impulse,
which can be defined as the time integral of a force. The
generalized non-smooth contact laws are expressed in terms of
impulse $\overrightarrow{P_C}$ and formal relative velocity
$\overrightarrow{\bar V_C}$ at contact point C. If $V_{CN}^-$,
$V_{CT}^-$, $V_{CN}^ +$ and $V_{CT}^ +$ denote the normal and
tangential relative velocities at contact point respectively
before and after collision, the formal normal and tangential
relative velocities are defined as follows:

\begin{eqnarray}
\left \{ \begin{array}{cccc} \bar V_{CN}  & = & \frac{V_{CN}^+ +
e_N V_{CN}^-} {1 + e_N}, \\
\\
\bar V_{CT}  & = & \frac{V_{CT}^+ + e_T V_{CT}^-} {1 + e_T },
\end{array} \right.
\end{eqnarray}

\noindent where $e_N$ and $e_T$ measure the inelasticity of
collisions and reduce to the normal and tangential restitution
coefficients in the case of binary collisions.

These generalized contact laws support momentum propagation
through contact networks inherent to dense assemblies of
grains. For a given time step, impulses and velocities are
determined according to an iterative process using a non-linear
Gauss-Seidel like method ~\cite{Jean99}. 
{In the case of large size packings of rigid grains, the CD method supports larger time steps, leading potentially to faster calculations than the molecular dynamic method for which small time steps are needed.}

\begin{figure}[htbp]
\includegraphics*[width=1.0\columnwidth]{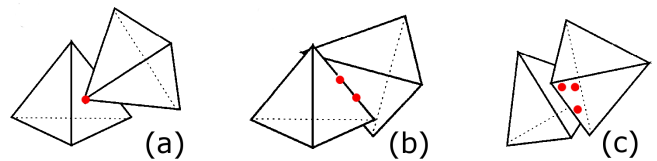}
\caption{(Color online) Two polyhedra can experience simple contacts (a), double contacts (b) or triple contacts (c). \label{situations-contact}}
\end{figure}

The CD method was applied using the LMGC90\copyright \ 
  platform
~\cite{Dubois03,Radjai11} which namely implements a 3D
extension of a 2D contact detection algorithm described with
details in~\cite{Azema_PRE_2007}. Basically, contacts with a given grain
are sought exclusively among its neighbors. When a neighbor is
located closer to the grain than a threshold distance called
\emph{gap}, a 3D extension of the \emph{shadow-overlap method}
devised by Moreau~\cite{Saussine06,Dubois03} is applied. In case
of overlap between the grain shades, their contact plane is
determined. Four contact situations may be encountered (Fig.
~\ref{situations-contact}): vertex-to-face or edge-to-edge, represented by a
single point and called \emph{simple}; edge-to-face, represented
by two points and called \emph{double}; finally, face-to-face,
represented by three points and called \emph{triple}
(vertex-to-edge and vertex-to-vertex being very unlikely to
happen). These situations allow identifying a contact plane
and compute the contact impulse and velocity components at each contact point.

This method proved apt to deal with dense flows of disks~\cite{Chevoir01a,Lois_PRE_2005,Lois_EPL_2006} as well as with quasi-static plastic deformation~\cite{Azema_PRE_2006,Azema_PRE_2007,Estrada_PRE_2008,Azema_EPJE_2008,Azema_MechMat_2009,Azema_PRE_2010}.

\subsection{Materials and system parameters} \label{subsec:parameters}

The present study focuses on monodisperse sphere packings (MSP), bidisperse sphere packings (BSP) and monodisperse pinacoid packings (MPP).

The spacing of lateral walls $W$ takes discrete values between $5d_L$ and $20d_L$, and the sample period along the $x$ axis is $20d_L$. With the final height $h$ of the packing in the range of $16d_L$ to $20d_L$, the number $n$ of grains varies between 1,900 and 30,400 for spheres, depending on the proportion by volume of small grains, and between 3,600 and 15,000 for pinacoids.

The time step value $\Delta t$ is taken small enough to moderate the grain interpenetration 
{incumbent to grains displacement between two successive implementations of the contact detection algorithm,} but sufficiently large to keep the calculation duration reasonable. 
{In this perspective,} limiting to $d_L/100$ the translation of grains during $\Delta t$ {seems appropriate}. For our grain packings subject to compaction under their own weight, this leads to the following relation:

\be \label{eq.I} \frac{d_L}{100} = v_{max}.\Delta t, \ee

\noindent with $v_{max} = \sqrt{2g\Delta h}$ the maximum speed reached by a grain free falling from initial height of the deposited packing down to the altitude of the packing free surface at the end of the compaction. Hence, equation~\ref{eq.I} leads to the following expression for the time step:

\be \label{eq.II} \Delta t = \frac{1}{100}.\sqrt{\frac{d_L}{2\Delta h}}.\sqrt{\frac{d_L}{g}}. \ee

Although the between-grain friction is set to zero, that of wall-grain contacts ($\mu_w$) is assigned non-zero values in a few simulations to study the influence of wall friction.
~~\\

The simulated system parameters are summarized in table~\ref{tab:sphere_packings} for spheres and in table~\ref{tab:pina_packings} for pinacoids. They are expressed as dimensionless quantities by defining the following normalization terms: lengths and times
are respectively measured in units of $d_L$ and $t_0=\sqrt{d_L/g}$, the characteristic free fall time of a rigid grain of diameter $d_L$ subject to gravity exclusively. For a given set of system parameters, three grain packings 
{are} simulated in order to average the various measured quantities.

\begin{table}[htbp]
\begin{center}
\scriptsize
  \caption{Sphere packings}
  \begin{tabular}{c|c|c|c|c|c|c|c}
    \hline
    \hline
    $n$ & $x_s~(\%)$ & $L_x/d_L$ & $L_y/d_L$ & $h/d_L$ & $\Delta t/\sqrt{d_L/g}$ & $\mu_w$ & $e_n$,$e_t$\\
    \hline
    \hline
    $2300$ & $0$ & $20$ & $5$ & $16$ & $3.10^{-4}$ & $0.0$ & $0.0$\\
    $~\rm{to}$ & $10$ &  & $10$  & $~\rm{to}$ & $~\rm{to}$ & $0.3$ & \\
    $30400$ & $25$ &  & $20$ & $20$ & $10^{-3}$ & $0.5$ & \\
     & $40$ &  & &  &  & $1.0$ & \\
    \hline
    \end{tabular}
  \label{tab:sphere_packings}
\normalsize
\end{center}
\end{table}

\begin{table}[htbp]
\begin{center}
\scriptsize
  \caption{Pinacoid packings}
  \begin{tabular}{c|c|c|c|c|c|c}
    \hline
    \hline
    $n$ & $L_x/d$ & $L_y/d_L$ & $h/d_L$ & $\Delta t/\sqrt{d_L/g}$ & $\mu_w$ & $e_n$,$e_t$\\
    \hline
    \hline
    $3600$ & $20$ & $5$ & $16$ & $3.10^{-4}$ & $0.0$ & $0.0$\\
    $~\rm{to}$ &  & $10$ & $~\rm{to}$ & $~\rm{to}$ &  & \\
    $15000$ &  & $20$ & $20$ & $10^{-3}$ &  & \\
    \hline
  \end{tabular}
  \label{tab:pina_packings}
\normalsize
\end{center}
\end{table}

\section{State of packings}\label{sec:state_packing}

In order to examine the influence of wall-induced confinement on the solid fraction and structure of dense packings for various grain shape and polydispersity, it is necessary to adopt a reference packing state and to ensure that the compaction method used allows to approximate such a state while providing sufficient repeatability for a given set of materials and system parameters. As mentioned in paragraph~\ref{subsec:init-solicit}, the compaction method used consists in depositing rigid frictionless grains (with or without wall friction) under their own weight. For sphere packings with presumably negligible confinement, several authors have experimentally~\cite{macrae_61,emam_05} or numerically~\cite{Zhang2001,Silbert_PRE_2002,emam_05} observed that this compaction method led to random close-packed states characterized by the generally agreed solid fraction value of $0.64$. According to~\cite{roux_04}, random close-packed states of rigid frictionless grains (spherical or non-spherical) are equivalent to packing states in which the grains are homogeneously spread and in a stable equilibrium without crystallization or segregation (observe that the notion of "stable equilibrium" refers to the minimization of a potential energy that ensures maximum solid fraction~\cite{Roux_PRE_2000}). Furthermore, extensive investigation of the random close-packed state carried out by~\cite{Agnolin_PRE_2007} with spherical particles has evidenced the uniqueness of this state in the limit of infinitely large samples subject to fast isotropic compression (to avoid cristallization). Hence, the influence of wall-induced confinement on the solid fraction and structure of dense packings may be assessed against the random close-packed state taken as the reference. Keeping in mind that our compaction method only allows to approximate the random close-packed state (our compression is not isotropic) and that the uniqueness of this reference state has only been evidenced {for} sphere packings, it is expected that meeting as much as possible the criteria stated by~\cite{roux_04} will lead to sufficiently repeatable solid fraction and microstructure characteristics for a given set of materials and system parameters to observe confinement effects for various grain shapes and polydispersity. Therefore, preliminary assessment consists in checking the state of our simulated packings (both sphere and pinacoid packings) in terms of stable equilibrium, homogeneity and reasonable interpenetration given the particularities of the contact dynamics method. Further assessment will be undertaken in sections~\ref{sec:solid_fraction} and~\ref{sec:packing-microstructure}.

\subsection{Equilibrium}

According to~\cite{Agnolin_PRE_2007}, grain packings for which the following criteria are met on each grain have reached a stable equilibrium:

\be \label{eq.III} \sum F < 10^{-4}d^2P, \ee 
\be \label{eq.IV} \sum M < 10^{-4}d^3P, \ee
\be \label{eq.V} E_c < 10^{-8}d^3P. \ee 

\noindent where $\sum F$, $\sum M$ and $E_c$ are respectively the net force, net momentum and total kinetic energy of the grain. Indeed, ~\cite{Agnolin_PRE_2007} have observed that setting to zero all grain velocities in such a state and letting the packing relax further did not lead to any kinetic energy or unbalanced force level regain beyond these threshold values.
~~\\

As a consequence, these criteria were used to check the attainment of a stable equilibrium state in our simulations, which was the case for all of them.

\subsection{Interpenetration}\label{subsec:interpene}

The grain interpenetration, calculated as the sum of interpenetrated volumes between neighboring grains divided by the sum of grain volumes, was checked in the bulk 
{region} of the packing at the end of each simulation. 

For sphere packings, the interpenetration was calculated analytically as the sum of interpenetrated volumes between couples of spheres (for a given couple of spheres, the interpenetrated zone consists of two spherical caps) and it was found to be very low (in the range of $10^{-5}$ to $10^{-3}\, \%$ by volume).

For pinacoids, a routine was designed to compute the solid fraction as well as lower and upper bounds 
of the grain interpenetration. Basically, this routine consists of superimposing a lattice on the grain packing and calculating the solid volume in each cell of the lattice. For a given cell, this solid volume is the sum of elementary volumes analytically calculated from the intersection between any pinacoid and the cell. 
\begin{figure}[htbp]
\begin{center}
		\includegraphics[width=1.0\columnwidth]{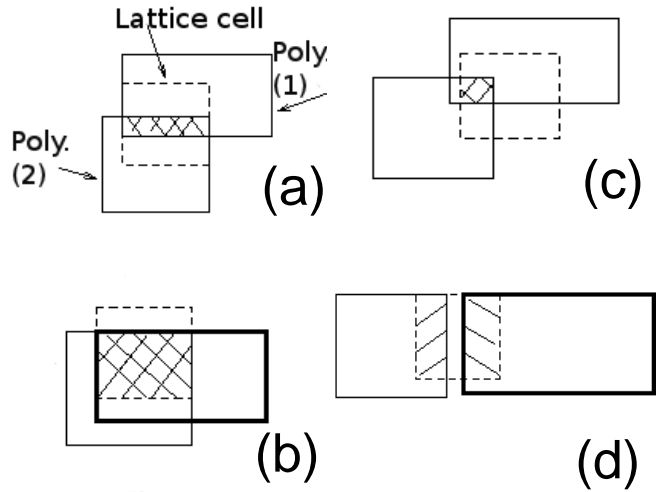} 
\caption{ 2D simplified representation of the four intersection situations between two pinacoids and a lattice cell: (a) solid volume $\geq$ cell volume and partial interpenetration in the cell; (b) solid volume $\geq$ cell volume and total interpenetration in the cell; (c) solid volume $<$ cell volume and interpenetration (partial as represented or total); (d) solid volume $<$ cell volume and no interpenetration.} \label{fig:interpene}
\end{center}
\end{figure}
In order to bound the grain interpenetration, one shall focus on cells intersected by two neighboring pinacoids, leading to one of the four situations depicted on Fig.~\ref{fig:interpene} (in 2D for simplicity reasons): 
\begin{itemize}
\item In situations $a$ and $b$, the solid volume $V_{sol}$ contained by the cell is in excess of actual cell volume $V_{cell}$, hence the lower bound of actual interpenetrated volume is $(V_{sol} - V_{cell})/2$ (situation $a$) whereas the upper bound is $V_{sol}/2$ (situation $b$);
\item In situations $c$ and $d$, the solid volume contained by the cell is smaller than actual cell volume, hence the lower bound of actual interpenetrated volume is $0$ (situation $d$) whereas the upper bound is $V_{sol}/2$ (situation $c$).
\end{itemize}
Observe that $V_{sol}/2$ is the upper bound of the interpenetrated volume no matter the situation. When their size decrease, the lattice cells that are intersected by two pinacoids tend to concentrate exclusively inside actual interpenetrated areas ($IA$) where $V_{sol}/2 = V_{cell}$ or astride their border ($AB$) where $0 < V_{sol}/2 < V_{cell}$. Hence, the total interpenetrated volume of the packing $V_I$ is bounded by the following interval:\\
\\
$V_I \in [\sum \limits_{cell \in IA} V_{cell}\,;\sum \limits_{cell \in IA} V_{cell}+\sum \limits_{cell \in AB} V_{cell}]$,
\bigskip

\noindent in which $\sum \limits_{cell \in AB} V_{cell}$ tends to zero with decreasing cell size.
 
\bigskip
For each pinacoid packing geometry, table~\ref{tab:pina_interpen} gathers lower $I_{min}$ and upper $I_{max}$ bounds of grain interpenetration, e.g bounds of $V_I$ expressed as a percentage of the packing solid volume. These values were computed in the bulk 
{region} using a lattice with $d_L/20$-large cubical cells, and each of them was averaged over three simulations. The interpenetration calculated in our pinacoid packings, in the range of 3 to 5\% by volume, is clearly much higher than the one calculated for sphere packings. 
\begin{table}[htbp]
\begin{center}
  \caption{Pinacoid packings interpenetration for MPP.}
  \begin{tabular}{c c c c c}
    \hline
    \hline
 $W/d_L$& $5$& $10$& $20$ & $PBC$\\
    \hline
 $I_{min}$ (\% vol.)& $4.6$& $4.5$& $3.4$ & $3.5$\\
    \hline
 $I_{max}$ (\% vol.)& $5.0$& $4.7$& $3.6$ & $3.7$\\
     \hline
  \end{tabular}\label{tab:pina_interpen}
  \end{center}
\end{table}

A first reason to explain these differences lies with the determination of contact between two grains. In the case of sphere packings, this determination is very simple and requires no interpenetration: grains are in contact when the distance between their centers is lower or equal to the sum of their radii. Such a contact is only one point, which is located on the segment connecting the centers of spheres at a distance of each sphere center equal to its radius. Besides, the orientation of the contact normal is borne by the segment connecting the grain centers. In the case of pinacoid packings, the determination of contact between two grains is much more complex, time-consuming and implies more or less interpenetration: first, grains are in contact when their respective shadows always overlap no matter the projection direction. Hence, much more calculation than for sphere packings shall be performed to prove the existence of a contact, and the simultaneous achievement of these overlap situations generally implies some interpenetration. Next, in case of a contact, it may not be a unique point but rather two (edge-to-face contact) or three (face-to-face contact) points as explained in section~\ref{subsec:NSCD}, which obviously leads to more interpenetration. 

A second reason lies with the non-smooth approach associated with the contact dynamics. In molecular dynamics~\cite{Frenkel2001}, contact forces increase proportionally to a power function of the interpenetration, leading to high repulsion contact forces thus low interpenetration in the limit of rigid grains. In the contact dynamics method where no such relation is applicable, the interpenetration is 
{namely} monitored by the quality of the convergence of impulses and velocities at contact points within the range of permissible values on the generalized non-smooth contact laws. Hence, {in addition to an appropriate time step value,} a low level of interpenetration requires optimizing both the convergence criteria and number of Gauss-Seidel iterations while keeping the calculation time acceptable (for more information, refer to~\cite{Moreau94,Moreau96,Dubois03,Radjai11}). 

Anyhow, the contact dynamics method is known to give interpenetration values in the order of a few percent by volume {(see~\cite{Saussine_04})}, and our quest of the densest possible disordered packing made of frictionless particles unsurprisingly leads to interpenetration values in the higher range. Hence the interpenetration evidenced by our results is acceptable.

\subsection{Homogeneity of distribution}

\begin{figure}[htbp]
\begin{center}
\includegraphics*[width=\columnwidth]{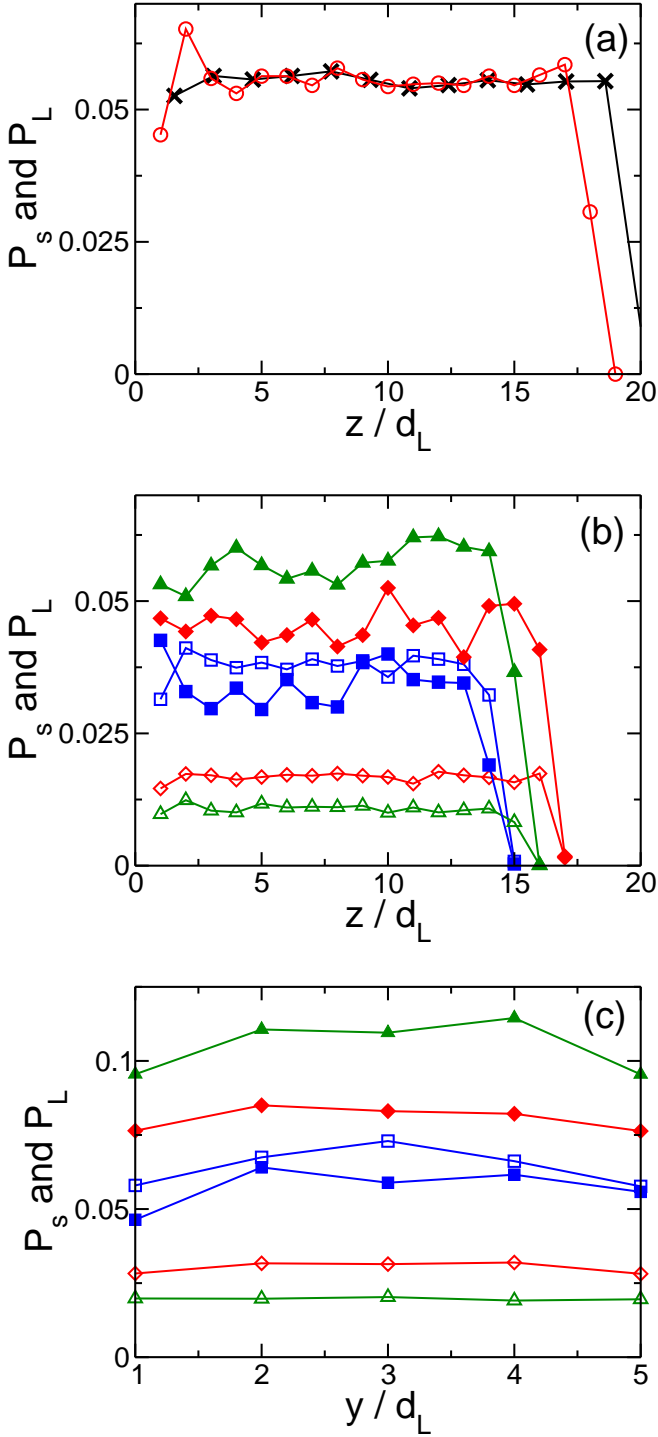}
\caption{(Color online) (a) Homogeneity of MSP ($\circ$) and MPP ($\times$) along the $z$ axis, expressed as the ratio of the number of grains to the total number of grains in $d_L$-thick layers.
({b}) Homogeneity of BSP along the $z$ axis, expressed as the ratio of the number of large (empty symbols) and small (full symbols) grains to the total number of grains in $d_L$-thick layers for $x_s=10\%$ ($\square$), $x_s=25\%$ ($\diamond$) and $x_s=40\%$ ($\triangle$). (c) Homogeneity of BSP along the $y$ axis, same calculation method, same key. For (a), {(b)} and (c), the gap between sidewalls is $W=5d_L$ and the data have been averaged over three simulations.}\label{fig:phi_z}
\end{center}
\end{figure}

In order to ensure that the applied compaction method leads to homogeneously distributed packings, we examine the variations in the proportions of large ($P_L$) and small ($P_S$) grains with distance $z$ from the bottom wall {(Fig.~\ref{fig:phi_z} (a) and (b))}. Therefore we count the number of particles of each size in $d_L$-thick regions of the packing and divide that number by the total number of grains. 
{Although small deviations (that may be due to segregation) close to the bottom of our packings are observed, the proportion profiles are almost constant showing that grains in sphere or pinacoid packings are reasonably vertically homogeneously distributed.} 
{The absence of segregation along the $y$ axis is also checked for BSP in the homogeneous zone (e.g. far from bottom and free surface). Fig.~\ref{fig:phi_z} (c) displays the variations of proportions $P_L$ and $P_s$ in $d_L$-thick layers parallel to the sidewalls. The proportion profiles are almost constant, showing reasonable horizontal homogeneity.}

\section{Solid fraction}\label{sec:solid_fraction}
\subsection{Average solid fraction}

In this subsection, our aim is to study the effect of confinement on the solid fraction of MSP, BSP and MPP, that is to say for various proportions of small particles and various grain shapes. For this purpose, we report the evolution of the aforementioned quantity for several values of gap between sidewalls. We will also test the {geometrical} model  mentioned in the introduction (cf Eq.~\ref{eqn:weeks}).
{The solid fraction is computed 
{from analytical calculation of the volume} of each sphere or each pinacoid present within a given volume. {This volume incorporates any particle located $3d_L$ away from the bottom wall and the free surface.} For the solid fraction of sphere packings,  the use of the Vorono\"i tessellation~\cite{Richard1999,Richard_EPJE_2001} gives the same results.}

\begin{figure}[htbp]
\begin{center}
\includegraphics*[width=1.0\columnwidth]{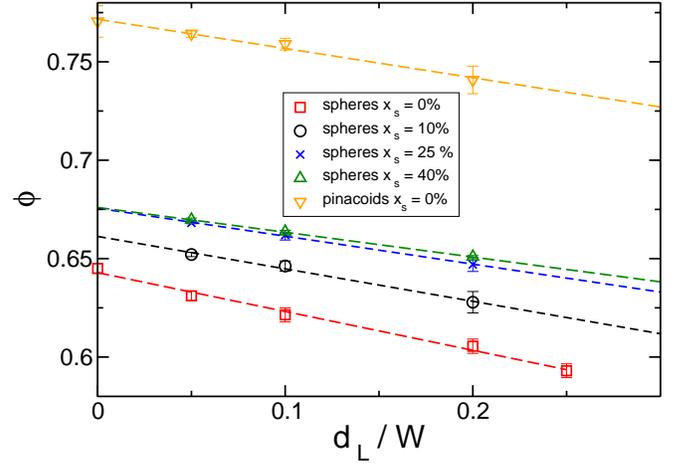}
\caption{(Color online) Plot of the average solid fraction versus $d_L/W$ for MSP, BSP and MPP. The lines are fits from the geometrical model
[Eq.~\ref{eqn:weeks}]. Error bars denote the standard deviation.}\label{fig:phi_vs_dsurW_MSP}
\end{center}
\end{figure}

Figure~\ref{fig:phi_vs_dsurW_MSP} reports the average solid fraction for BSP, MSP and MPP versus $d_L/W$. A first observation is that for a fixed $d_L/W$ value, an addition of small grains in a monosized sphere packing increases the solid fraction. This well-known phenomenon can easily be understood by considering two limit cases. The first one consists of a monosized sphere packing to which we add a few small particles ($x_s\ll 100\%$). In this case, small grains partially fill the porosity of the monosized packing and increase the solid fraction. The second limit case corresponds to a packing of small grains to which we add a few large particles ($x_s\approx 100\%$). The largest particles can then be considered as islands in a sea of small grains whose solid fraction is equal to that of a monosized packing: $\phi_{mono}$. Since the solid fraction of the islands is equal to $1$, the  average packing fraction is greater than $\phi_{mono}$. \\
More interestingly, an excellent agreement between our data and the {geometrical} model 
is found. The corresponding values of $C$ and $\phi_{bulk}$ are reported in table~\ref{tab:weeks}. It should be pointed out that the value of $\phi_{bulk}$ obtained for MSP is consistent with that of the random close packing ($0.64$) reported in the literature~\cite{Torquato2000}.
\begin{table}[htbp]
\begin{center}
\begin{tabular}{|c|c|c|c|c|c|}
\hline
\hline
 & \multicolumn{4}{|c|}{spheres} & pinacoids\\
 \hline
 \hline
$x_s$ & 0\% &10\% & 25\% & 40\%& 0\% \\
\hline
\hline
$C/d_L$&0.197&0.165&0.142 &0.126& 0.149  \\
\hline
$\Phi_{bulk}$& 0.643 & 0.661& 0.676 & 0.676 & 0.772\\
\hline
\end{tabular}
\caption{Values of $C$ and $\phi_{bulk}$ obtained by fitting the data reported in Fig.~\ref{fig:phi_vs_dsurW_MSP} with equation~(\ref{eqn:weeks}). 
}\label{tab:weeks}
\end{center}
\end{table}
 
Note that in Ref.~\cite{Desmond2009}, Desmond and Weeks compare {the geometrical model} with simulations of bidisperse sphere packing (50-50 binary mixture with particle size ratio of 1.4)  in the absence of gravity. Our results show that the validity of this model is much broader since it still holds in the presence of gravity for monodisperse sphere packings, for bidisperse sphere packings (independently of the fraction of small grains) as well as for monodisperse pinacoid packings. This result is important in the framework of real granular materials whose grains are far from being perfect spheres. 
Let us recall that the fit parameter $C$ is equal to $2h (\phi_{bulk}-\phi_{BL})$ (see section~\ref{sec:intro}). Our results show that when the fraction of small grains, $x_s$, increases, $C$ decreases. This can be the consequence of a decrease of the distance of propagation of the sidewall effects $h$ or/and of the difference $\phi_{Bulk}-\phi_{BL}$. 
To address this point we will study the local variation of the solid fraction close to the sidewalls. 
This is the objective of next subsection.\\


\subsection{Solid fraction profiles}\label{subsec:solid_fraction_profiles}
\begin{figure}[htbp]
\begin{center}
\includegraphics*[width=1.0\columnwidth]{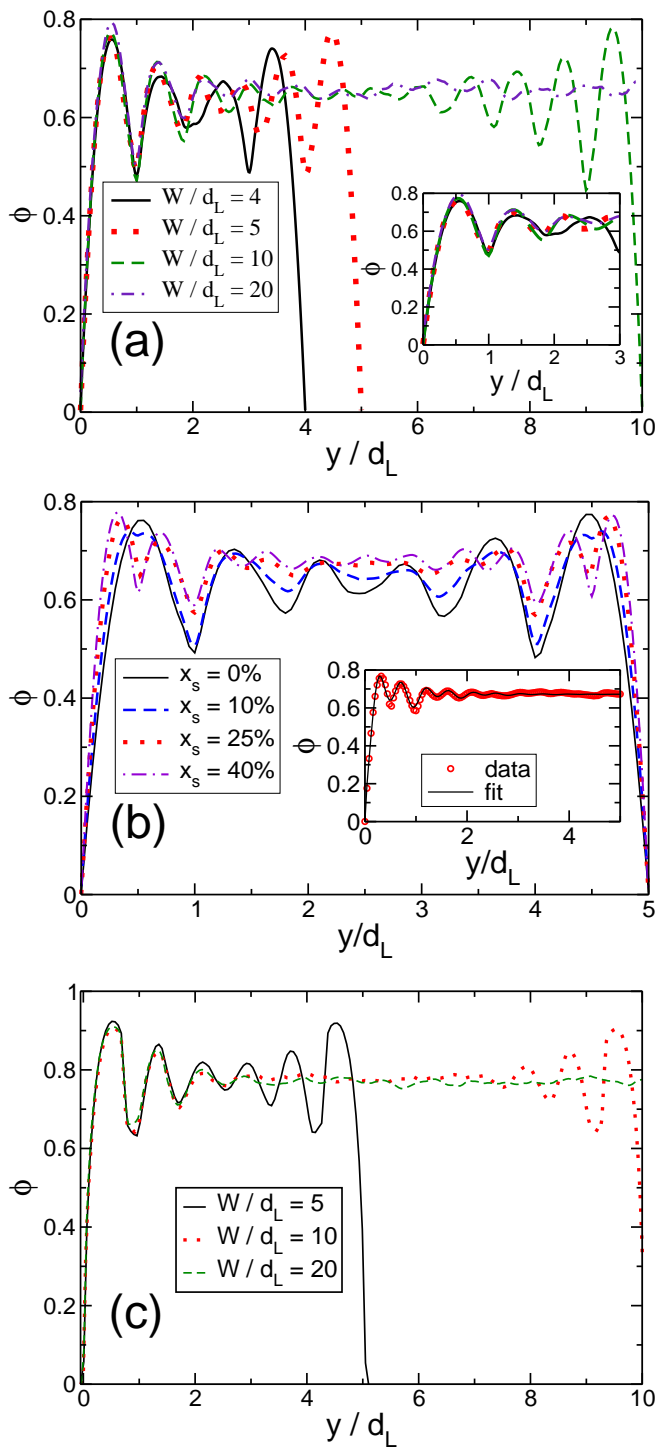}
\caption{(Color online) Solid fraction profiles as a function of distance $y/d_L$ from confining wall for 
(a) MSP with $W=4d_L$, $W=5d_L$, $W=10d_L$ and $W=20d_L$, (b) BSP with $W=5d_L$ and $x_s = 0\%, 10\%, 25\%\mbox{ and }40\%$,
and (c) MPP with $W=5d_L$, $W=10d_L$ and $W=20d_L$. Fluctuations of the local solid fraction are due to the layering of particles in the vicinity of the sidewalls. The inset in Fig.~\ref{fig:Suzuki}(a) is a zoom over $3d_L$. The inset in Fig.~\ref{fig:Suzuki}(b) shows the solid fraction for BSP (here $x_s=40\%$) and the corresponding fit (Eq.~\ref{eqn:fit_phi}.).
}\label{fig:Suzuki}
\end{center}
\end{figure}
Figure~\ref{fig:Suzuki} depicts the solid fraction profile as a function of the distance $y/d_L$ to the left sidewall for MSP (a), BSP (b) and MPP (c). The local solid fraction fluctuates with the distance from the wall, especially in the neighborhood of sidewalls and, if $W$ is large enough, it reaches a uniform value away from the sidewalls. 
The inset of Fig.~\ref{fig:Suzuki}(a) reports the packing fraction fluctuations as a function of the non-dimensional distance from the wall $y/d_L$.
{The aforementioned fluctuations 
{clearly reflect the} layering due to the presence of sidewalls
-- \textit{i.e.} an order propagation in the $y$ direction
~\cite{Suzuki2008}.}
For MSP, the confinement effect propagates over approximately $3d_L$ to $4d_L$. As a result, packings for which  $W<6d_L$ to $8d_L$ are influenced by the presence of walls over their full width. 
{In other words, for such size, the order generated by the sidewalls propagates in the whole packing. }
On the contrary for BSP as well as for MPP, the propagation seems to be shorter (approximately $1.5d_L$ to $2d_L$ for BSP and about $2d_L$ for MPP). 
{The presence of bidispersity or non-sphericity induces disorder 
{in} the vicinity of the sidewalls which 
{mitigates} the layering.}
To quantify more precisely the sidewall effects we have fitted the solid fraction profiles reported in Fig.~\ref{fig:Suzuki} by the following empirical law: 
\begin{align}
\label{eqn:fit_phi}
\phi(y)=&\left[1-\exp(-\frac y\alpha)\right]\times\nonumber\\
 &\left[\phi_{bulk}+\sum_{i\in\{S,L\}}\beta_i\cos\left(\frac{\pi(2y-1)}{\gamma_i}\right)\exp\left(-\frac y\lambda_i\right)\right]
\end{align}
 {In this expression, the characteristic lengths of the sidewall effect propagation for large {(L)} and small  {(S)} grains are respectively  $\lambda_L$ and $\lambda_S$. Parameter $\alpha$ characterizes the solid fraction increase close to the sidewalls, $\gamma_i$  {and $\beta_i$ respectively} 
correspond to the period  {and amplitude} of the structuration oscillations 
 {caused to} the solid fraction profile  {by the layering of small and large particles.} 
 {For monosized packings, we use the aforementioned fit with $\beta_S=0$}. An example of the obtained fits is plotted in the inset of Fig.~\ref{fig:Suzuki}b.}
Let us stress out that the fit used is purely empirical. Our aim is to obtain a reasonable approximation for confinement effect propagation rather than a precise description of the solid fraction profiles by an equation. 
\begin{figure}[htbp]
\begin{center}
\includegraphics*[width=1.0\columnwidth]{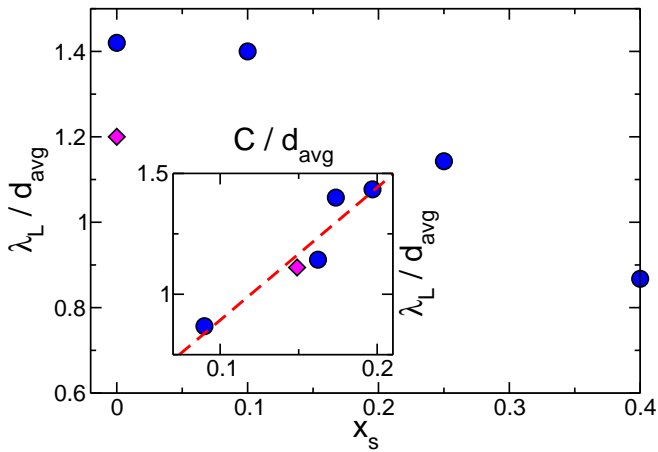}
\caption{(Color online) Characteristic length of confinement effect $\lambda_L$ versus fraction of small grains $x_s$ for MSP and BSP (circle) as well as for MPP (diamond) with $W=10d$. That length is defined in equation~\ref{eqn:fit_phi}. The effect of confinement is found to decrease with increasing grain polydispersity or grain angularity. The inset reports the same length versus that of the fit parameter in eq.~(\ref{eqn:weeks}): $C=2h(\phi_{bulk}-\Phi_{BL})$. The dashed line corresponds to a linear fit.}\label{fig:lambda}
\end{center}
\end{figure}

{The values of $\lambda_L$ (those of $\lambda_S$ are not statistically relevant for $x_S < 0.25$), normalized by the average grain size $d_{avg}$}, obtained this way are reported in Fig.~\ref{fig:lambda} for $W=10d_L$.
For sphere packings, the {normalized} characteristic length is found to decrease when the fraction of small spheres $x_s$ increases. Indeed, for $x_s=0$ we have $\lambda_L/d_{avg}\approx 1.4$ whereas $\lambda_L/d_{avg}\approx 0.85$ for $x_s=40\%$. This decrease proves that the polidispersity mitigates {the} confinement effect. 
{Morover, the fact that 
$\lambda_L/d_{avg}$ decreases with $x_s$ demonstrates that 
{$\lambda_L$ decreases quicker than the mean grain size.}}
For MPP we obtain $\lambda/ d_{avg} = 1.2$ which is smaller than the value obtained for MSP. This indicates that the sidewall effect is also 
{mitigated by an increase in grain angularity}.
Hence, 
characteristic length $\lambda_{{L}}$ is expected to correlate with the 
 {thickness} $h$  {of the boundary layers} introduced 
  {in equation~\ref{eqn:weeks}.} 
 In the inset of Fig.~\ref{fig:lambda} we report $\lambda_{{L}}$ versus $C{=2h(\phi_{bulk}-\phi_{BL})}$ and observe a good linear correlation between these two parameters. Furthermore, the data for both sphere and pinacoid packings 
 collapse on the same straight line whose intercept is equal to zero.\\

\subsection{Effect of grain/wall friction}
{So far, the presented simulations were performed with frictionless grains and sidewalls. However, additional simulations were performed to investigate the influence of grain/wall friction. For this purpose, the friction coefficient between grains was kept equal to zero, whereas the grain/wall friction coefficient $\mu_{gw}$ was successively set to $0.3$, $0.5$ and $1$. As before, three grain packings were simulated for each grain/wall friction coefficient in order to average the measured quantities.} {Our aim is not 
{to} address this point, but just to mention that in our contact dynamic simulations,  
we found that the grain/wall friction had no effect since neither the average solid fraction nor the solid fraction profiles were affected by $\mu_{gw}$. This result demonstrates that the influence of confinement on packing fraction is purely geometrical.}


\section{Packing microstructure}\label{sec:packing-microstructure}

Section~\ref{sec:state_packing} has established that our packings are homogeneous and that they have reached a stable equilibrium with acceptable interpenetration. {Then, in section~\ref{sec:solid_fraction} we have 
{verified}} that the simulated compaction method 
{allows to accurately achieve} the $0.64$ solid fraction characteristic of the random close-packed state of 
{monodisperse sphere packings when PBC are substituted for sidewalls.} Moreover this method is sufficiently repeatable to show significant influence of the confinement on the solid fraction of various grain packings. Now, section~\ref{sec:packing-microstructure} focuses on the internal state of our packings in order to investigate the influence of confinement on their microstructure. 
We first investigate the presence of 
{textural} order (section~\ref{subsec:order}).
Then, we study a usual characteristic to describe the microstructure of grain packings: the mean number of contacts per grains (coordination number). For various grain shapes and polydispersities, section~\ref{subsec:coord_number} discusses the influence of confinement on that characteristic. 

\subsection{Order}\label{subsec:order}
In this section, our aim is to 
{investigate the presence} of long-range 
{textural} order in the packings. {Let us 
{point out} here that by long-range order we mean an order that extends to the system size when 
{this size tends towards infinity~\citep{Ricci_PRE_2006}.}}\\
{In a granular packing, textural order may take various forms~\cite{cambou_09}: 
%
 translational arrangements of grains that combine to form patterns, preferential orientation of the contact network, preferential orientation of non-spherical grains themselves. Each of these aspects is addressed in the following paragraphs.}

First, translational arrangements are studied by means of the pair correlation function $g(r)$~\citep{Silbert_PRE_2002}. 
\begin{figure}[htbp]
\begin{center}
{\includegraphics*[width=1.0\columnwidth]{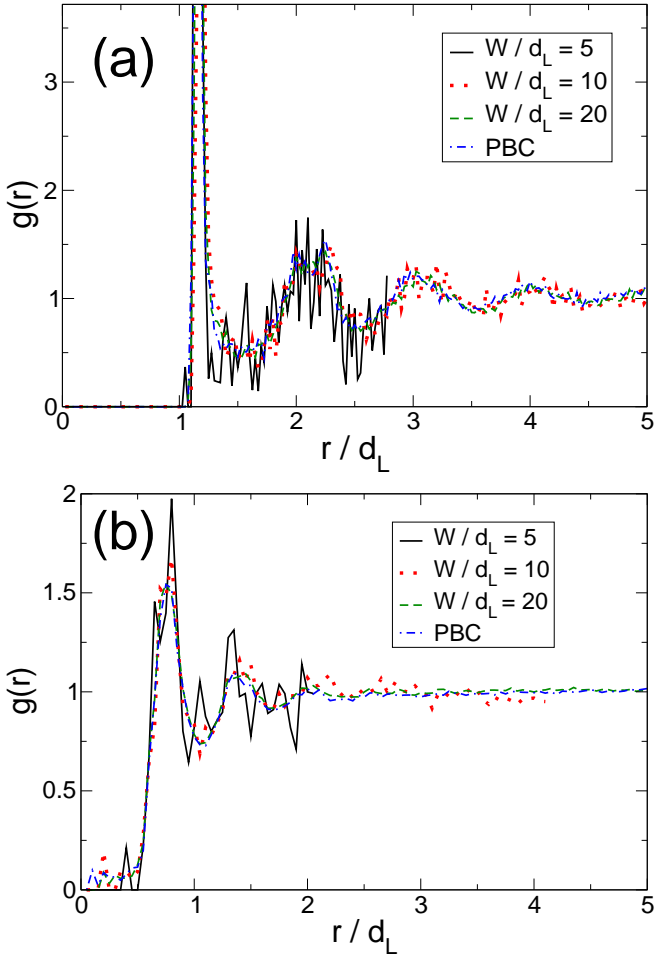}}
\caption{(Color online) Pair correlation functions of MSP (a) and MPP (b) for several values of  $W/d_L$. These exhibit local order that is stronger for MSP compared to MPP. When $r/d_L$ is large enough $g(r)$ tends towards one, indicating the absence of long range translational order.}\label{fig:gr}
\end{center}
\end{figure}
This function is depicted in Fig.~\ref{fig:gr} for MSP (a) and MPP packings (b) and for several values of $W$. 
{For both packings, local order extends over a few particle diameters, slightly less for MPP than for MSP due to the higher angularity of the former resulting in a loss of rotational symmetry. As a consequence, very confined packings exhibit ordering over their full size. However, for lower confinement (e.g. when $W/d_L = 10$ or more), $g(r)$ tends towards $1$ when $r$ increases beyond $3d_L$, indicating the absence of translational long range ordering within our packings.} 

\begin{figure}[htbp]
\begin{center}
\includegraphics*[width=1.0\columnwidth]{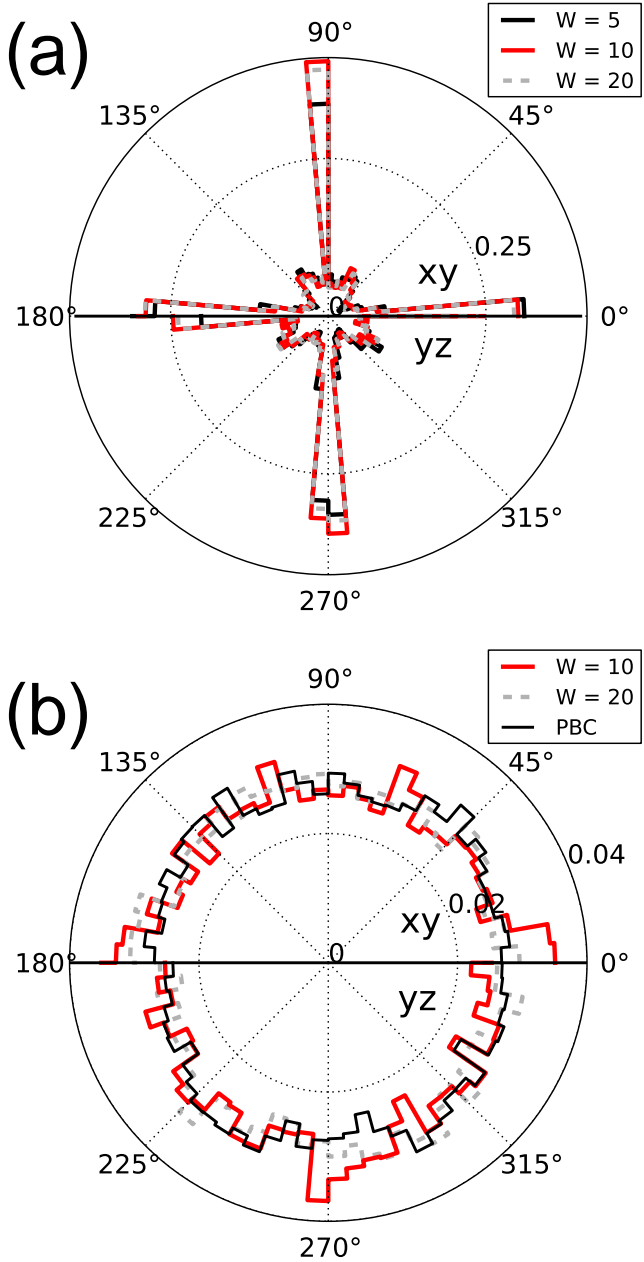}
\caption{(Color online) 
{Contact normal orientation distributions for MSP  inside the boundary layers 
 (a), 
  and inside the bulk region (b). 
 Upper half of each chart (e.g. from $0$ to $180^\circ$) corresponds to the $xy$ plane, while lower half corresponds to the $yz$ plane. Several gap widths are considered.}}\label{fig:contact-orient-MSP-W10-20-PBC}
\end{center}
\end{figure}
\begin{figure}[h!tbp]
\begin{center}
\includegraphics*[width=1.0\columnwidth]{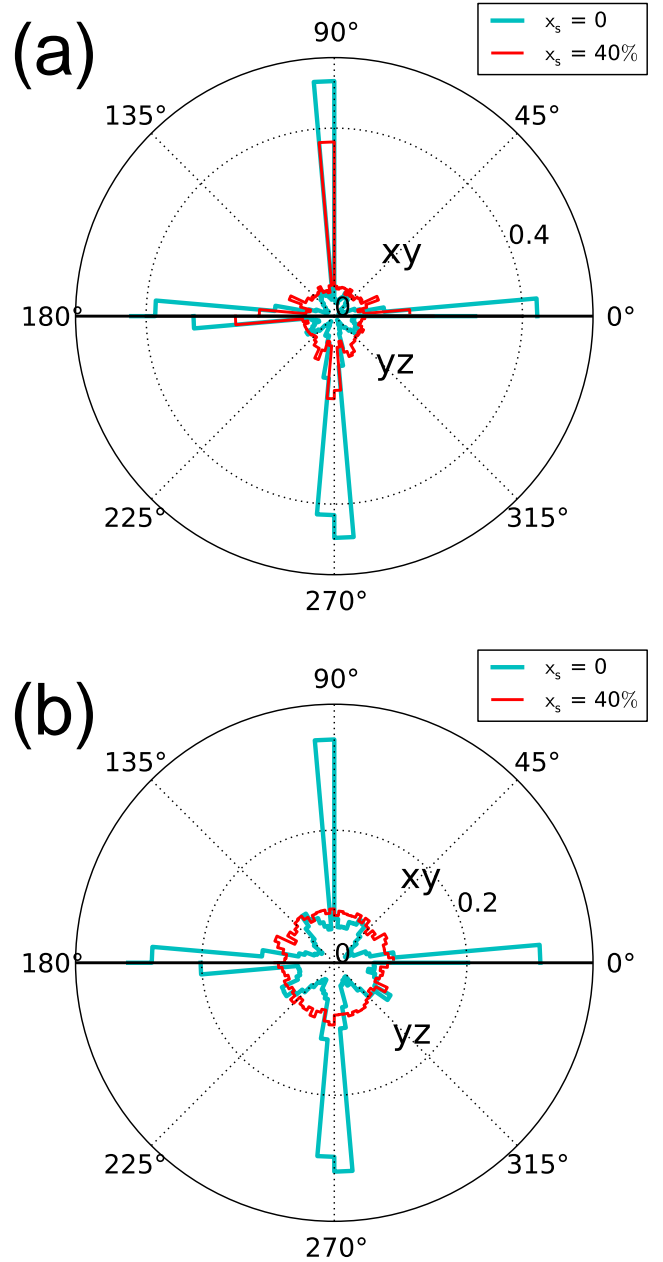}
\caption{(Color online) 
{Contact normal orientation distributions for MSP ($x_S=0$) and BSP ($x_S=40\%$) inside the boundary layers/closer than $1.5d_L$ to a sidewall 
 (a), 
 and inside the bulk region (b)
. Upper half of each chart (e.g. from $0$ to $180^\circ$) corresponds to the $xy$ plane, while lower half corresponds to the $yz$ plane. Gap width is $W=5d_L$ and bulk region coincides with particles located at least $1.5d_L$ away from sidewalls.}}\label{fig:contact-orient-MSP-BSP-W5}
\end{center}
\end{figure}
\begin{figure}[htbp]
\begin{center}
\includegraphics*[width=1.0\columnwidth]{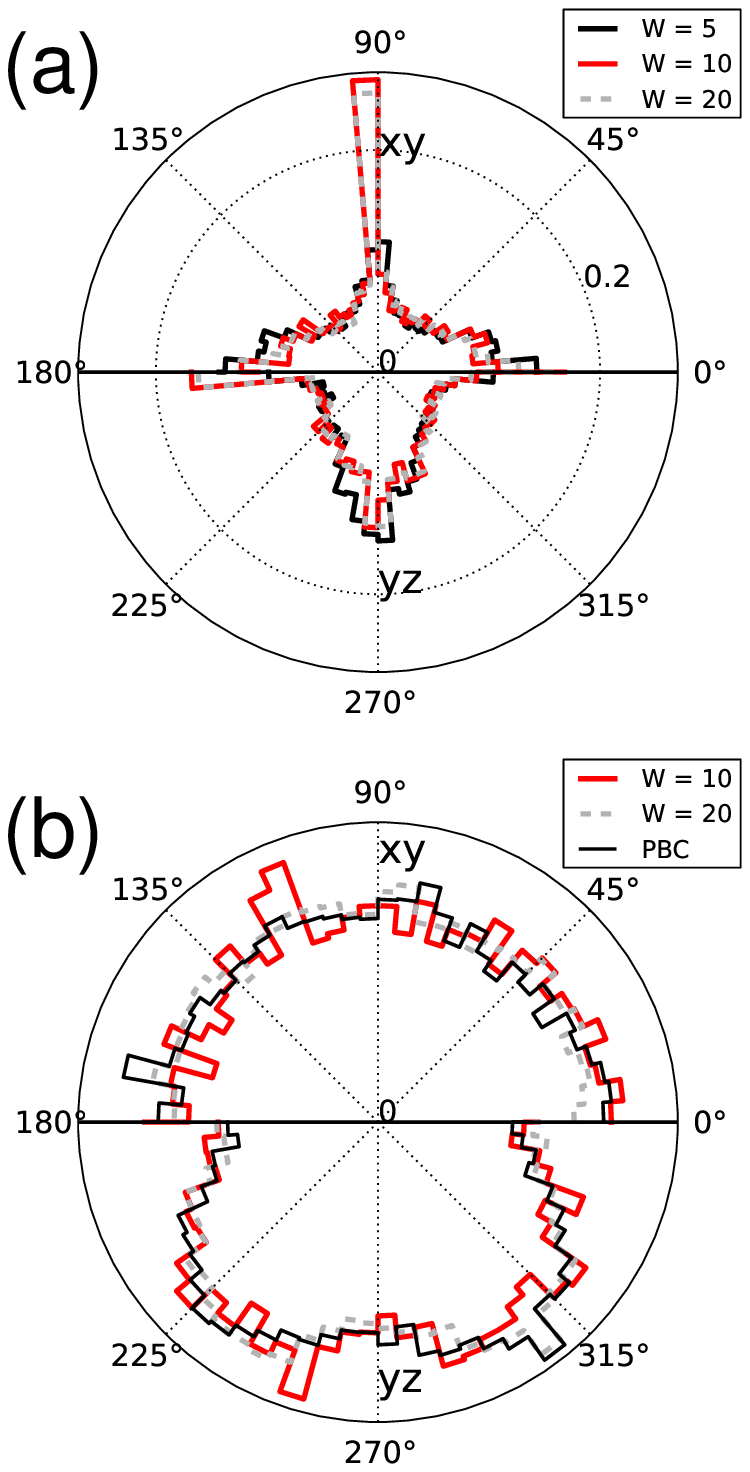}
\caption{(Color online) 
{Contact normal orientation distributions for MPP inside the boundary layers 
 (a), 
  and inside the bulk region (b). 
 Upper half of each chart (e.g. from $0$ to $180^\circ$) corresponds to the $xy$ plane, while lower half corresponds to the $yz$ plane. Several gap widths are considered.}}\label{fig:contact-orient-MPP-W10-20-PBC}
\end{center}
\end{figure}

Next, the existence of preferential orientations of the contact network is investigated. For this purpose, Fig.~\ref{fig:contact-orient-MSP-W10-20-PBC} displays for various confinements $2D$ representations of the distributions of contact orientations in MSP away from the bottom plane and the free surface. Given the sidewall-induced-layering evidenced in the $h$-thick boundary layers (see section~\ref{subsec:solid_fraction_profiles}), contact from the boundary layers (Fig.~\ref{fig:contact-orient-MSP-W10-20-PBC}a
) have been dealt with separately from those located in the bulk 
{region} (Fig.~\ref{fig:contact-orient-MSP-W10-20-PBC}b)
. Note that no bulk region is present in monodisperse packings where $W/d_L=5$ and, conversely, no boundary layer occurs in PBC packings. Furthermore, inside the boundary layers (see Fig.~\ref{fig:contact-orient-MSP-W10-20-PBC}a), an 
 anisotropy of contact orientations is visible regardless of the confinement in the $x$, $y$ and $z$ directions, 
 as well as at roughly $60^\circ$ to the $x$ direction in the $xy$ plane and $30^\circ$ to the $y$ direction in the $yz$ plane (corresponding to compact clusters of three spheres close to the sidewalls)
 . This anisotropy is fully consistent with the vertical layering of monodisperse packings close to the sidewalls, with a larger peak in the $y$ direction due to the high proportion of sidewall-sphere contacts. Unsurprisingly, in the bulk  {region}, Fig.~\ref{fig:contact-orient-MSP-W10-20-PBC}b 
 shows that the distribution of contact orientations remains isotropic for all these confinements. In order to assess the effect of polydispersity on the existence of preferential orientations of the contact network, 2D representations of the distributions of contact orientations are depicted in Fig.~\ref{fig:contact-orient-MSP-BSP-W5} for $W=5d_L$-thick packings, away from bottom plane and free surface. Observe that the substitution of $x_S=40\%$ by mass of small particles for large ones results in the emergence of a central $2d_L$-thick quasi-isotropic bulk region (see Fig.~\ref{fig:contact-orient-MSP-BSP-W5}b
). 
 Furthermore, note that the 
 boundary layers remain anisotropic (see Fig.~\ref{fig:contact-orient-MSP-BSP-W5}a) although the presence of small particles in between large ones tends to disturb the vertical layering of the latter (because the centers of inertia of small particles are not necessarily coplanar with those of large particles). 
  Hence, the anisotropy along the axes $x$, $y$ and $z$ is mitigated, 
  whilst other preferential orientations corresponding to various patterns made of small and large grains each in contact with the others are generated
 . Eventually, in order to assess the effect of grain shape on the existence of preferential orientations of the contact network, 2D representations of the distributions of contact orientations in MPP away from the bottom plane and the free surface are represented in Fig.~\ref{fig:contact-orient-MPP-W10-20-PBC} for various confinements. As for sphere packings, contacts located in boundary layers have been dealt with separately from those located in the bulk region. Inside the boundary layers, 
 an anisotropy of contact orientations is visible regardless of the confinement in the $x$, $y$ and $z$ directions 
 (see Fig.~\ref{fig:contact-orient-MPP-W10-20-PBC}a), and this anisotropy may be explained by the wall-induced layering just like for sphere packings. 
 In the bulk region, 
 Fig.~\ref{fig:contact-orient-MPP-W10-20-PBC}b 
 shows that pinacoid packings 
 exhibit isotropic contact orientation distributions in the $xy$ plane, 
 but not along the $z$ axis where, unlike for sphere packings, some anisotropy is visible even for packings with PBC. This anisotropy may be explained by the deposition under gravity protocol, with pinacoids rotating around their center of inertia under steric hindrance contraints in order to minimize their potential energy.

\begin{figure}[htbp]
\begin{center}
\includegraphics*[width=1.0\columnwidth]{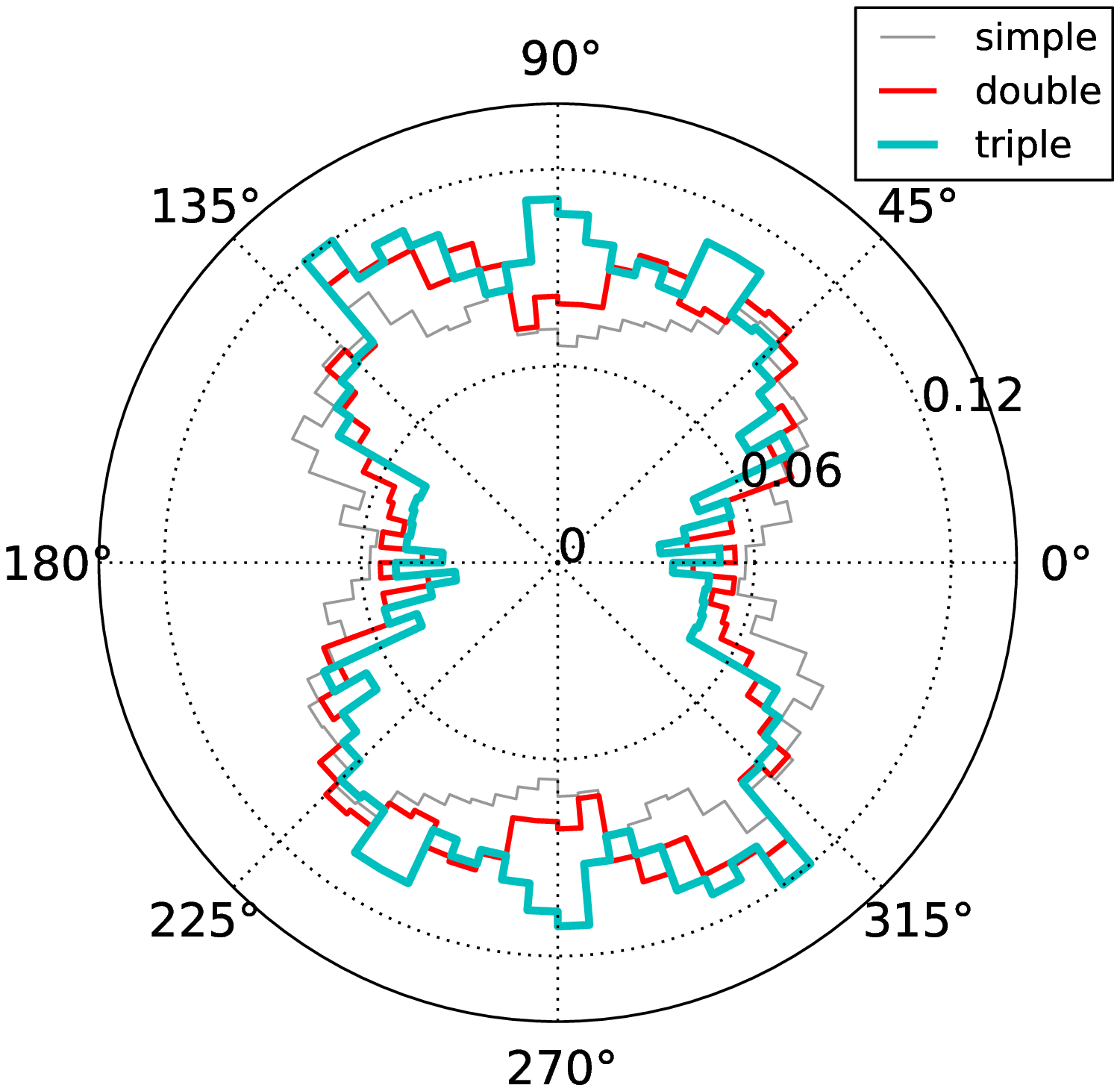}
\caption{(Color online) Distribution of the orientations of simple (a), double (b) and triple (c) contacts for MPP in the $yz$ plane. 
{Periodic Boundary Conditions are used in the $x$ and $y$ direction.}}\label{fig:distri_sdt}
\end{center}
\end{figure}

The anisotropy observed in Fig.~\ref{fig:contact-orient-MSP-BSP-W5}b 
 and~\ref{fig:contact-orient-MPP-W10-20-PBC}b 
 respectively for BSP with $x_S=40\%$ and MPP with PBC may be calculated and compared to that of the isotropic reference state depicted in Fig.~\ref{fig:contact-orient-MSP-W10-20-PBC}b 
 with PBC. Given the rotational symmetry of the contact normal orientation distributions in the $xy$ plane (see Fig.~\ref{fig:contact-orient-MSP-W10-20-PBC}b
 ,~\ref{fig:contact-orient-MSP-BSP-W5}b 
 and~\ref{fig:contact-orient-MPP-W10-20-PBC}b
 ), the anisotropy may be quantified using a second order development of the contact orientation probability density function $P(\vec{n})$ (see~\cite{Azema_MechMat_2009} for details):
\be \label{eq. proba_contact} P(\vec{n})=\frac{1}{4\pi}[1+a(3\cos^2\theta-1)], \ee
where:
\begin{itemize}
\item $a=5(F_3-F_1)/2$ denotes the branch vector coefficient of anisotropy derived from	eigenvalues $F_3$ and $F_1$ of the fabric tensor~\cite{cambou_09,Radjai11,Azema_MechMat_2009}
\item $\theta$ denotes the polar coordinate in the $xy$ plane
\end{itemize}
This coefficient 
may vary from $0$ (perfectly isotropic packing) to $2.5$ (perfectly anisotropic packing).
Table~\ref{tab:anisotropy} gathers values of the branch vector coefficient of anisotropy calculated for MSP with PBC, in the bulk region of BSP with $x_S=40\%$ and for MPP with PBC. These values show no significant differences between the bulk region of BSP with $x_S=40\%$ and MSP with PBC. Furthermore, the coefficient of anisotropy of branch vectors obtained for MPP with PBC remains below $0.1$, denoting a rather small 
 anisotropy. Last, observe in Fig.~\ref{fig:distri_sdt} showing the distributions of orientations of simple (face-vertex), double (face-edge) and triple (face-face) contacts that the vertical anisotropy in MPP with PBC is identical no matter the contact type. As a consequence, no long-range contact orientation anisotropy is generated in our frictionless grain packings by the grain deposition protocol used, except a weak anisotropy generated in pinacoid packings along the $z$ axis.

\begin{table}
  \caption{
  {Branch vector coefficient of anisotropy calculated for MSP with PBC ($W/d_{L}=20$), in the bulk {region} of BSP with $x_S=40\%$ ($W/d_{L}=5$) and for MPP with PBC ($W/d_{L}=20$).}}
  \begin{tabular}{c c c c}
    \hline
    \hline
 $Configuration$ & $MSP~(PBC)$& $BSP~(W/d_{L}=5)$ & $MPP(PBC)$\\
    \hline
$a$ & $0.028\pm0.011$ & $0.032\pm0.005$ & $0.081\pm0.032$\\
    \hline
    \hline
  \end{tabular}
  \label{tab:anisotropy}
\end{table}

Finally, the orientations of particles that are not symmetric by rotation may also be a source of anisotropy within the packing. 
To detect 
{a preferential orientation of pinacoids in such a packing, one may} use the nematic order parameter $Q_{00}^2$
. Here, we recall briefly how this parameter can be determined 
 (for details, refer to~\cite{Camp_JChemPhys_1997,John_JChemPhys_2004}). 
For each particle, if we call $\overrightarrow{u},\overrightarrow{v},\overrightarrow{w}$ the 
 {unit vectors} of its base of inertia (which, in our case, 
 {align with} its axes of symmetry) we can define the following tensor~\cite{John_JChemPhys_2004} :
\[Q^{uu}_{\alpha\beta}=\frac{1}{n}\sum_{i=1}^{n}\left(\frac{3}{2}u_{i\alpha}u_{i\beta}-\frac{1}{2}\delta_{\alpha\beta}\right)\;\;\;\text{with}\;\;\;\alpha,\beta=1,2,3\]
and where $n$ is the number of particles and $\delta$ the Kronecker symbol. We apply the same definition with $Q^{vv}_{\alpha\beta}$ and $Q^{ww}_{\alpha\beta}$.
From those tensors, the nematic order parameter can be determined:
\[Q_{00}^2={}^t\overrightarrow{Z}.Q^{zz}.\overrightarrow{Z},\]
where $\overrightarrow{Z}$ is the eigenvector corresponding to the 
{larger} eigenvalue of the three tensors $Q^{uu}$, $Q^{vv}$ and $Q^{ww}$. $Q^{zz}$ is the corresponding diagonalized tensor.
By construction, this parameter varies between $0$ and $1$. 
{For each of our pinacoid packings, two values of the nematic order parameter have been calculated, one corresponding to particles located in a boundary layer, and one corresponding to particles located in the bulk {region}. Whatever the confinement, the nematic order parameter ranges between $0.05$ and $0.07$ in the bulk {region}, which is quite low and shows the absence of privileged grain orientation, whereas it is slightly higher in the boundary layers (between $0.105$ and $0.128$).}

As a conclusion of this subsection, 
frictionless 
{grain} packings used in the present work do not exhibit significant long range order,  
except a weak anisotropy of the contact orientation distributions 
observed in pinacoid packings along the $z$ axis. 
 Furthermore, sidewalls induce order close to their location that, in the case 
 of very confined packing, propagates over the whole system.

\subsection{Coordination number} \label{subsec:coord_number}

\begin{figure}[htbp]
\begin{center}
\includegraphics*[width=1.0\columnwidth]{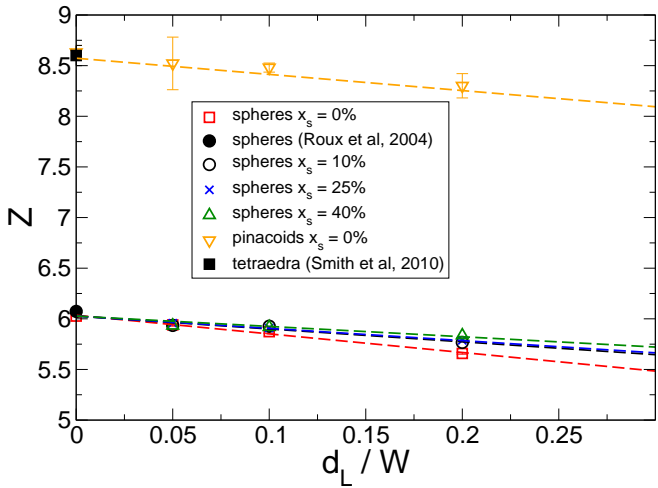}
\caption{(Color on-line) Coordination number as a function of $d_L/W$ for MSP and MPP. Error bars on pinacoid packings results denote the standard deviation (not represented for sphere packings because errors are smaller than symbol size). 
The linear relationship between $Z$ and $d_L/W$ suggests that the {geometrical model}, initially derived for the packing fraction, is also valid for the coordination number.}\label{fig:weeks_coord}
\end{center}
\end{figure}

Figure~\ref{fig:weeks_coord} shows the variations of the coordination number with $d_L/W$ for MSP, BSP and MPP. 
Each value is averaged over three simulations and the error bars denote the corresponding standard deviation. Preliminary examination of our results obtained with bi-periodic boundary conditions (unconfined state with $d_L/W \to 0$) suggests the following remarks: for sphere packings, the calculated coordination number is $6.027\pm{0.012}$, which is very close to the $6.073\pm{0.004}$ value calculated by~\cite{roux_04} in the RCP state. For pinacoid packings, the calculated coordination number is $8.581\pm{0.068}$. Although no study of pinacoid packings could be found in the literature, such a high coordination number value has already been observed in disordered packings of particles having a similar shape ($8.6\pm{0.1}$ calculated by~\cite{Smith_PRE_2010} for packings of tetraedra upon extrapolation to the jamming point).\\
When confinement increases, the coordination number decreases linearly for both MSP and MPP, which is consistent with the linear decrease of the solid fraction evidenced in Fig.~\ref{fig:phi_vs_dsurW_MSP}. Though, MPP coordination number values tend to be more scattered than MSP ones, which could be due to a combination of finite packing size effects with the higher level of interpenetration observed in pinacoid packings.
Nevertheless, the aforementioned linear relation between $Z$ and $1/W$ suggests a generalization 
of {the}  {geometrical} model to the coordination number. For this purpose, let us define $Z_{bulk}$ and $Z_{BL}$, respectively the coordination number for the bulk region and the coordination number for the boundary layers. By writing the coordination number as the average of $Z_{bulk}$ and $Z_{BL}$ weighted by the thicknesses of their respective zones (resp. $W-2h_Z$ and $2h_Z$) we obtain
\begin{equation}
Z=\frac{W-2h_Z}{W}Z_{bulk}+\frac{2h_Z}{W}Z_{BL}=Z_{bulk}-\frac{C_Z}{W},
\end{equation}  
with $C_Z=2h_Z(Z_{bulk}-Z_{BL})$.

Figure~\ref{fig:weeks_coord} also shows that the influence of polydispersity on packing coordination number $Z$ decreases to zero when the confinement diminishes, which is consistent with~\cite{roux_05}. Indeed, in the unconfined state, the lack of contacts of small spheres  {with others (due to the steric hindrance of large ones)} 
 is compensated by the excess of contacts of large spheres  {with small ones.}

\begin{figure}[htbp]
\begin{center}
\includegraphics*[width=1.0\columnwidth]{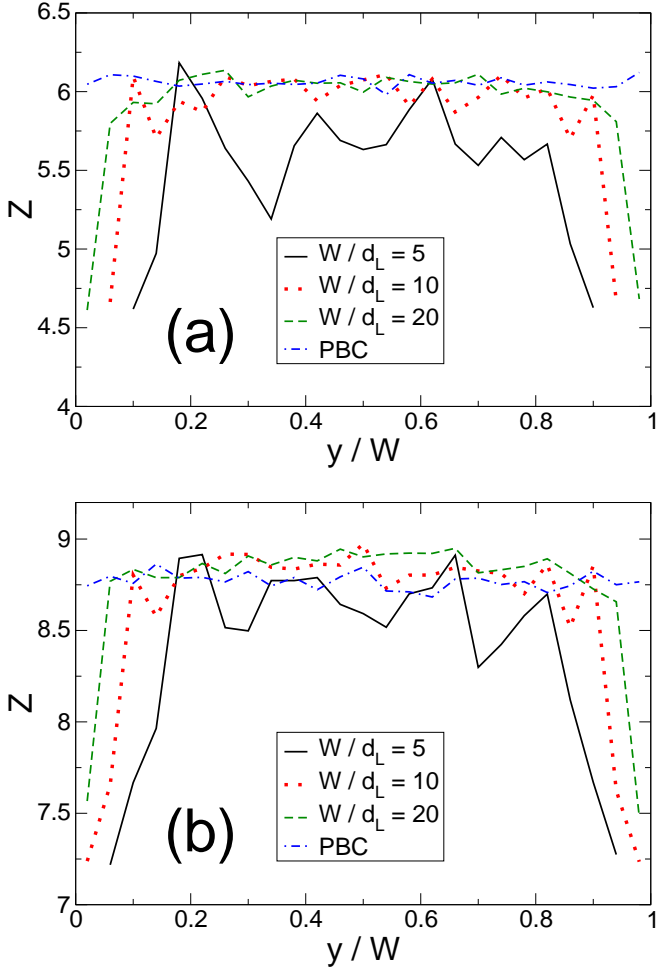}
\caption{(Color online) Coordination number profiles (along $y$) for MSP (a) and for MPP (b) for several gap width. These profiles evidence a constant central zone and two drop zones in contact with the sidewalls.}\label{fig:prof-Z}
\end{center}
\end{figure}

To investigate the coordination number decrease with increasing confinement, Fig.~\ref{fig:prof-Z} depicts coordination number profiles in the $y$ direction (normal to the sidewalls) for sphere and for pinacoid packings. 
Each of these profiles is averaged over three simulations and is determined upon subdividing the packing into  slices perpendicular to the $y$ direction and calculating for each slice the average number of contacts per particle having its center of inertia in the slice. In confined state, all these profiles evidence a central zone where the coordination number is almost unchanged compared to the unconfined reference state (except for sphere packings with $W=5d_L$), and two "drop zones" in contact with the sidewalls where the coordination number symmetrically drops by $1.4$ (for sphere packings) to $1.5$ contacts (for pinacoid packings) from their respective unconfined reference state. The thicknesses of these drop zones look identical to that of the boundary layers described 
in  {the}   {geometrical model}~\cite{Desmond2009}, leading to the same conclusion that grain angularity mitigates the effect of 
 sidewalls  {on the coordination number drop in their vicinity}.\\
To confirm this observation, we may consider the {geometrical} model and compare 
$\zeta_\phi=C/\phi_{bulk}$ with $\zeta_Z=C_Z/Z_{bulk}$. For MSP, we obtain 
$\zeta_\phi=0.304$ and $\zeta_Z=0.305$ and for MPP $\zeta_\phi=0.186$ and $\zeta_Z=0.193$. 
Note that the values of $\zeta_\phi$ and $\zeta_Z$ are also comparable in the case of BSP.
The strong correlation between those two quantities shows that the propagation of the confinement effect  is comparable for the two studied quantities: $\phi$ and $Z$. 
As described in subsection~\ref{subsec:NSCD}, pinacoid packings incorporate simple, double and triple contacts and it is of interest to investigate the effect of confinement on their respective distribution.
\begin{figure}[htbp]
\begin{center}
\includegraphics*[width=1.0\columnwidth]{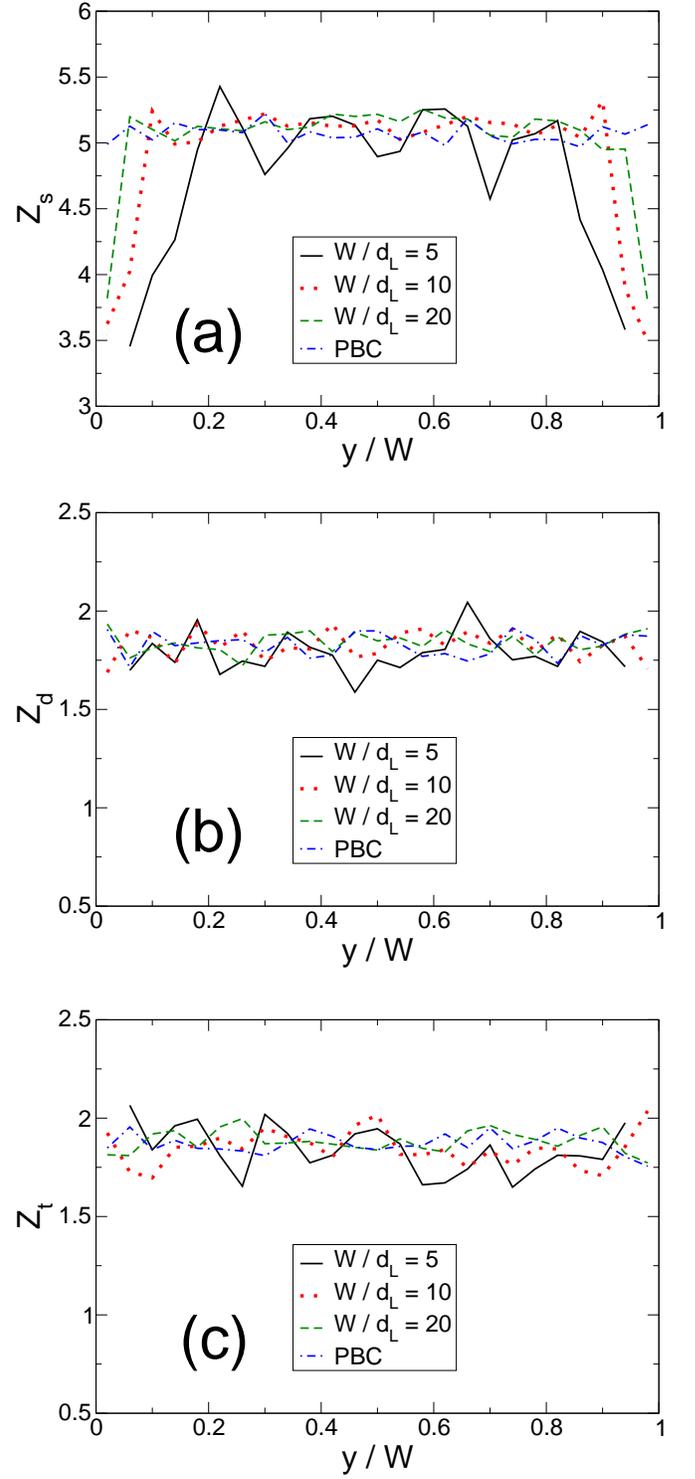}
\caption{(Color online) Coordination number profiles (along $y$) of MPP for simple (a), double (b) and triple contacts (c) for confined  ($W \in [5d_L,20d_L]$) and unconfined packings.}\label{fig:prof-ZSDT-MPP}
\end{center}
\end{figure} 
Therefore, Fig.~\ref{fig:prof-ZSDT-MPP} depicts the coordination number profiles of MPP for simple (a), double (b) and triple contacts (c). Like in Fig.~\ref{fig:prof-Z}, all these profiles evidence a central zone where coordination number values are almost unchanged compared to the unconfined reference state (except for sphere packings with $W=5d_L$). These values are $Z_s\approx 5$, $Z_d\approx 1.7$ and $Z_t\approx 1.8$ respectively for simple, double and triple contacts.

In order to check the relevance of these coordination number values, one shall observe that packings of $n$ frictionless rigid grains at equilibrium obey the following relation~\cite{Roux_PRE_2000} between the degree of hypostaticity $k_0$, the degree of hyperstaticity $h_0$, the number of contacts that carry forces $N_c = \frac{n}{2}.(Z_s + 2.Z_d + 3.Z_t)$, and the number of degrees of freedom $N_f = 6n$ of the packing:

\begin{align}
  \label{eqn:hyper}
   & 		& N_f+h_0=N_c + k_0 \nonumber\\ 
   & \Leftrightarrow	& 12 + 2.\frac{h_0}{n}= Z_s + 2.Z_d + 3.Z_t + 2.\frac{k_0}{n}
\end{align}

 {If we assume that} 
 the pinacoids in our packings are randomly oriented, 
 which seems reasonable according to the values of the nematic order parameter (see section~\ref{subsec:order}), {then} 
 no motion is possible without generating work in the contacts network, which means that the degree of indeterminacy of contact forces in the packing is zero, therefore $k_0$ should be set to $0$ in equation~\ref{eqn:hyper}. Upon incorporating in equation~\ref{eqn:hyper} the aforementioned coordination number values as well as that of $k_0$, one obtains:

\begin{equation}
  \label{eqn:hyper2}
   2.\frac{h_0}{n}= 13.8 - 12 = 1.8
\end{equation}

Observe that, for isostatic pinacoid packings (e.g. $h_0=0$), equation~\ref{eqn:hyper} would lead to $Z_s + 2.Z_d + 3.Z_t = 12$. Here, it is clear that $Z_s + 2.Z_d + 3.Z_t > 12$. The level of interpenetretation calculated in subsection~\ref{subsec:interpene} together with the finite size of packings may lead to a sum $Z_s + 2.Z_d + 3.Z_t$ slightly higher that $12$, but it is doubtful that this sum would reach $13.8$ upon this sole explanation. The presence of hyperstaticity in our pinacoid packings seems more realistic and at least consistent with equation~\ref{eqn:hyper2} and with our finding of as much as 2 triple contacts per grain ($Z_t\approx 1.8$).  
 Although interesting, further investigation of 
 the presence of hyperstaticity falls beyond the scope of the present paper.

Coming back to Fig.~\ref{fig:prof-ZSDT-MPP}, the profiles show that MPP exhibit more simple contacts than the sum of double and triple contacts. They also evidence that confinement primarily impacts the simple contact profiles, whereas double and triple contact profiles remain unchanged. As a consequence, the vicinity of sidewalls is not a privileged location for edge-to-face or face-to-face contacts, although a drop in the simple contact profiles tends to make them look over-represented.



\begin{figure}[htbp]
\begin{center}
\includegraphics*[width=1.0\columnwidth]{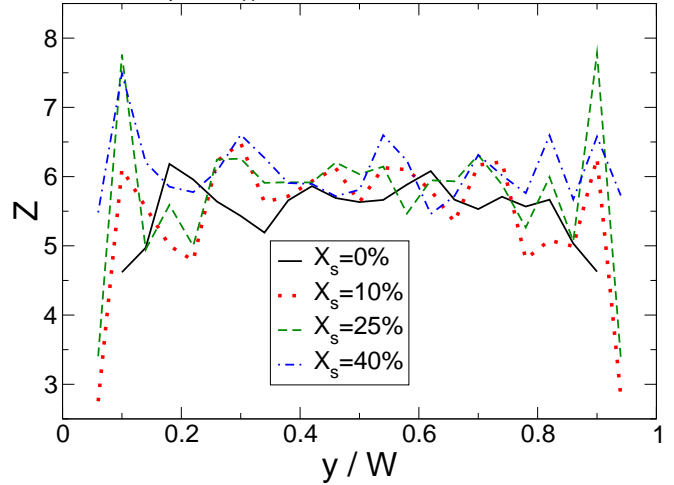}
\caption{(Color online) Coordination number profiles (with $y$) for BSP. The gap between sidewalls is $W=5d_L$.}\label{fig:profil_Z_BSP}
\end{center}
\end{figure}

For a fixed confinement, Fig.~\ref{fig:profil_Z_BSP} shows the influence of polydispersity on coordination number profiles in the $y$ direction (normal to the sidewalls) for sphere 
 packings. As before, each of these profiles is averaged over three simulations and is determined upon slicing the packing perpendicular to the $y$ direction. As observed in Fig.~\ref{fig:Suzuki}b, an increasing polydispersity does not seem to impact the bulk {region}, but rather reduces the thickness of the boundary layers, hence mitigates the effect of sidewalls confinement on the coordination number.

Finally, like for the solid fraction, we have not observed any effect of the grain-wall friction coefficient on the coordination number of MSP.\\

\section{Conclusion}\label{sec:conclusion}
In this work, we have shown how a confining boundary alters the solid fraction as well as the internal structure of static {frictionless}
granular materials {compacted under their own weight using}  
 the non-smooth
 contact dynamics {simulation} method. We did not restrict ourselves to sphere packings but extended our work to packings made of a particular type of polyhedra: pinacoids.\\
{As previously reported, the presence of sidewalls induces short-range order in their vicinity.} {Except a weak contact orientation anisotropy observed with pinacoid packings in the vertical direction, no long-range order was observed in our packings}. 
We have demonstrated that both the polydispersity and the angularity of grains lower the confinement effect. This effect has been observed for the solid fraction and for the coordination number.
Our results have shown that the  {geometrical model ~\cite{Verman_Nature_1946,Brown_Nature_1946,Combe_PhD_2001,Desmond2009}} that captures the linear evolution of the solid fraction versus $-1/W$  is valid for sphere packings as well as for pinacoid packings and that it holds whatever the packing polydispersity.\\
Interestingly, this model, initially derived for the packing fraction can be extended to capture the effect of confinement on the coordination number. The characteristic length quantifying the effect of the sidewalls is found to be the same for those two quantities.\\
Finally, we have shown that {the} effect of 
 wall friction is 
 negligible, indicating that the major influence of the confining sidewalls is geometric.\\
Several perspectives arise from this study, among which the need to address with more details the presence 
  {of 
  } hyperstaticity 
 {in} our packings of frictionless pinacoids.
\section*{Acknowledgments}
We thank F. Chevoir, J.-N. Roux, G. Saussine, N. Roquet and N. N. Medvedev for helpful conversations. Many thanks to the LMGC90 team in Montpellier for making their simulation platform freely available. We {are also} grateful to O. Garcin for technical support.

\end{document}